\documentclass{article}
\usepackage{booktabs}
\usepackage{spconf,amsmath,graphicx}
\usepackage{amsfonts}
\usepackage{arydshln}
\usepackage{cite}
\usepackage{mathptmx} 
\usepackage[T1]{fontenc} 
\usepackage{times} 

\usepackage{enumitem}
\setlist{nosep, leftmargin=14pt}

\usepackage{mwe} 

\title{Contrast-Free Myocardial Scar Segmentation in Cine MRI using Motion and Texture Fusion}
\name{\begin{tabular}{c}
Guang Yang$^{1,\star}$, Jingkun Chen$^{1,\star}$, Xicheng Sheng$^{5}$, Shan Yang$^{6}$ \\ 
Xiahai Zhuang$^{5}$, Betty Raman$^{2}$, Lei Li$^{1, 3, 4}$, Vicente Grau$^{1}$ \thanks{* Co-first authors with equal contributions.}
\end{tabular}}

\address{
    $^{1}$Department of Engineering Science, University of Oxford, Oxford, UK \\
    $^{2}$Radcliffe Department of Medicine, University of Oxford, Oxford, UK \\
    $^{3}$School of Electronics \& Computer Science, University of Southampton, Southampton, UK \\
    $^{4}$Department of Biomedical Engineering, National University of Singapore, Singapore \\
    $^{5}$School of Data Science, Fudan University, Shanghai, China\\
    $^{6}$ Department of Radiology, Zhongshan Hospital, Fudan University, Shanghai, China
}

\begin{document}
%
\maketitle

\begin{abstract}
Late gadolinium enhancement MRI (LGE MRI) is the gold standard for the detection of myocardial scars for post myocardial infarction (MI). 
LGE MRI requires the injection of a contrast agent, which carries potential side effects and increases scanning time and patient discomfort. 
To address these issues, we propose a novel framework that combines cardiac motion observed in cine MRI with image texture information to segment the myocardium and scar tissue in the left ventricle.
Cardiac motion tracking can be formulated as a full cardiac image cycle registration problem, which can be solved via deep neural networks.
Experimental results prove that the proposed method can achieve scar segmentation based on non-contrasted cine images with comparable accuracy to LGE MRI. 
This demonstrates its potential as an alternative to contrast-enhanced techniques for scar detection.
\end{abstract}


\section{Introduction}
\label{sec:intro}

Cardiovascular disease (CVD) remains the leading cause of premature mortality and morbidity worldwide, with myocardial infarction (MI) being one of the most prevalent cardiac conditions \cite{zeidan2024myocardial}. MI arises from the restriction of blood flow, leading to damage of myocardial (Myo) tissue and the eventual formation of a scar in the myocardium. Severe MI can cause heart failure, arrhythmia, and diminished contractility. Late gadolinium enhancement magnetic resonance imaging (LGE MRI) serves as the gold standard for left ventricle (LV) scar detection and quantification \cite{li2023myops}. Traditionally, scar tissue is analyzed either manually by experienced physicians or through semi-automated methods such as the $n$-standard deviation ($n$-SD) \cite{bondarenko2005standardizing} or full width at half maximum (FWHM) \cite{grani2019comparison}, which involve human-in-the-loop checking.

Deep learning (DL) approaches have been proposed for automatic myocardial scar segmentation \cite{li2022medical, chen2022semi, chen2024dynamic}. Most existing methods rely on LGE MRI, which requires contrast agent injection, posing potential side effects and significantly increased scan times and patient discomfort. Recent studies have explored scar analysis using contrast-free imaging as a more cost-effective alternative.
One prominent strategy involves the use of generative models, which aim to synthesize LGE-style images, allowing LGE-based analyses without contrast agents. 
For instance, Xu \textit{et al.} \cite{xu2020contrast} proposed sequential causal generative models that embed synthesis and scar segmentation tasks into adversarial learning, and Zhang \textit{et al.} \cite{zhang2021toward} combined cine with T1 mapping imaging to generate synthesized LGE images for scar segmentation. However these approaches do not provide a direct quantification of the scar and the prediction performance can heavily depend on the quality of synthesized LGE images. Another solution is to directly analyze cine MRI, leveraging its temporal information to detect abnormalities, such as in Xu \textit{et al.} \cite{xu2018direct} and Zhang \textit{et al.} \cite{zhang2019deep}. 
Additionally, Xu \textit{et al.} \cite{xu2020segmentation} directly estimated scars from cine MRIs without explicit cardiac motion extraction.
Instead, they employed the shared representations between image segmentation and quantification tasks (with infarct-related information) to assist the scar localization. 
Nonetheless, implicit motion extraction lacks interpretability, while most explicit approaches rely on optical flow applied to detect motion only between adjacent frames \cite{alfarano2024estimating}.
Other recent studies explored using ECG data to assist scar localization from cine MRI, though these methods have so far only been evaluated on simulated data \cite{li2024towards,lian2024frequency}.

In this work, we propose a novel cardiac motion and texture informed Myo scar segmentation (MTI-MyoScarSeg) model based on cine MRI.
Our approach involves extracting cardiac motion by calculating the displacement fields from each frame relative to the end-diastolic (ED) phase. These motion features are then fused with the original image to integrate both motion and texture information, enabling accurate segmentation of LV Myo and scars.
\begin{figure*}[t]
    \centering
    \includegraphics[width=0.8\textwidth]{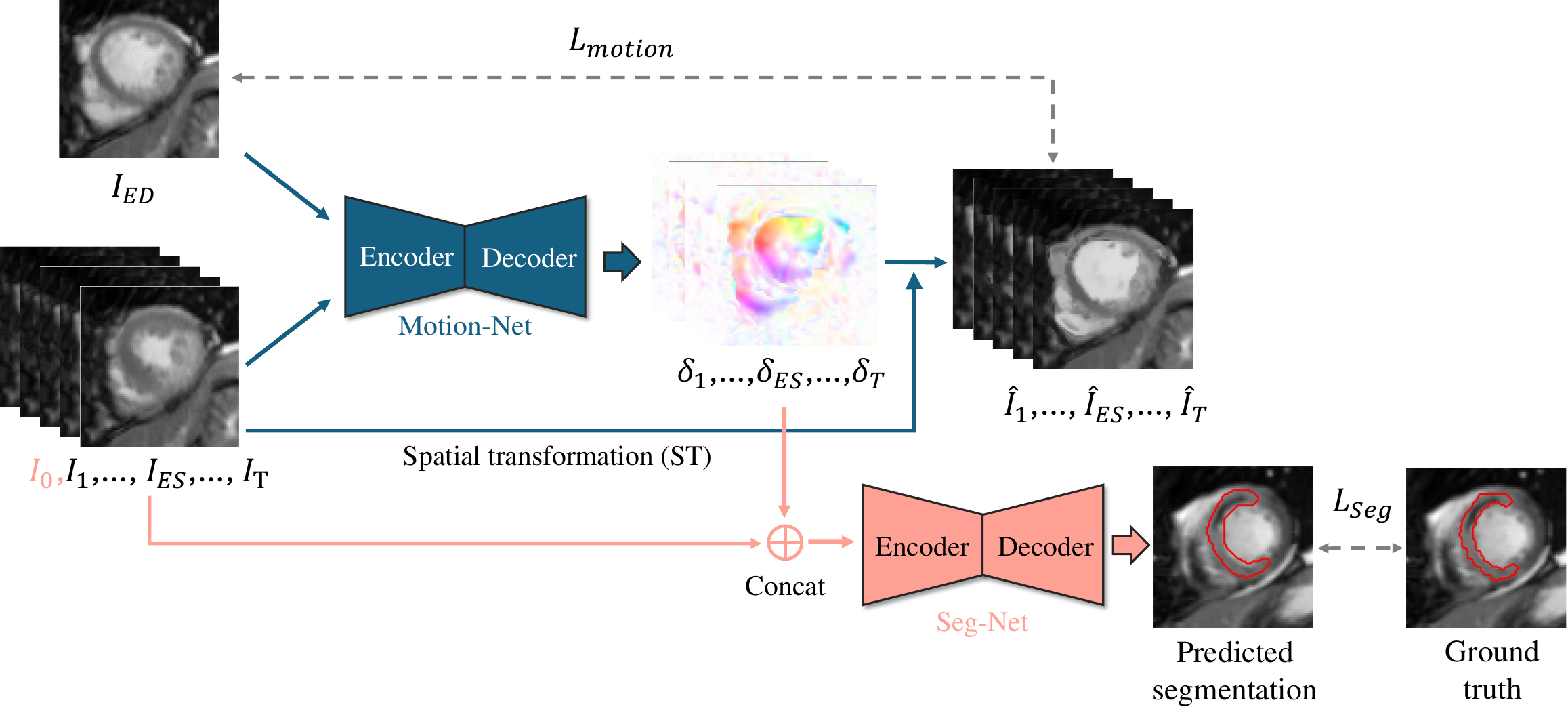}
    \caption{The architecture of the proposed cardiac motion and texture informed model for contrast-free myocardial scar localization from cine MRI. Note that the input of Seg-Net is the whole cardiac cycle (including $I_0$) and estimated cardiac motion.}
    \label{fig:flow_chart}
    \vspace{-4mm}
\end{figure*}
The main contributions of this work are:

\begin{itemize}
    \item We propose a novel contrast-free Myo scar segmentation framework that fuses motion and texture information from cine MRI sequences.
    \item We compare different motion extraction strategies and prove the effectiveness of using ED phase as a fixed reference frame.
    \item We demonstrate that our method can achieve performance comparable to LGE-based methods.
\end{itemize}

\section{Methods}
\label{sec:method}

The proposed MTI-MyoScarSeg model, illustrated in Fig.~\ref{fig:flow_chart}, comprises two key components: a motion extraction network (described in Section \ref{ssec:me}) and a Myo and scar segmentation network based on motion and texture information (detailed in Section \ref{ssec:seg}). The framework estimates cardiac motion throughout the cardiac cycle and integrates this motion information with the image sequences to enable precise segmentation of both the LV Myo and scars.

\subsection{Cardiac Motion Extraction}
\label{ssec:me}

MTI-MyoScarSeg model first employ a cardiac motion extraction network ($\mathit{F_{ME}}$) to process a sequence of cine MRI frames $I_t = \{I_0, ..., I_{T}\}$, where $I_0$ represents the ED phase and $I_{T}$ is the last frame of the sequence. 
The ED frame ($I_0$ or $I_{ED}$) serves as the reference frame, and the cardiac motion extraction network estimates corresponding displacement of each frame $I_t$, where $t \in \{1,...,T\}$, relative to $I_0$.

The motion extraction network adopts a U-Net architecture \cite{ronneberger2015u} combined with a spatial transformer (ST) \cite{jaderberg2015spatial}, allowing for the transformation of other frames into the ED reference frame. The model takes paired inputs of $I_t$ and $I_0$, predicting the optical flow field $\delta = (\delta_x, \delta_y)$, which represents the displacement in the $x$ and $y$ axes, respectively:
\begin{equation}\label{equ:1}
    \delta_t = \mathit{F_{ME}}(I_0, I_t),
\end{equation}
where $\delta \in \mathbb{R}^{H \times W \times 2}$. After obtaining the predicted displacement $\delta_t$, the corresponding frame $I_t$ is warped to align with the ED frame using the following warping operation:
\begin{equation}
    \hat{I_t} = \mathcal{T}(I_t, \delta_t),
\end{equation}
where $\mathcal{T}$ is a transformation function that aligns moving image $I_t$ by optical flow $\delta_t$ to fixed image $I_0$. This operation adjusts pixel positions based on the displacement field, so that for each pixel $(x, y)$, $\hat{I_t}(x,y) = I_t(x+\delta_x, y+\delta_y)$.

The following loss functions are minimized. First, the mean squared error (MSE) is used to maximize the similarity between the fixed ED image $I_0$ and the warped frames $\hat{I_t}$:
\begin{equation}
    \mathcal{L}_{\text{Motion}} = -\sum_{t=1}^{T} \text{MSE}(I_{0}, \hat{I_t}).
\end{equation}
Additionally, a smoothness loss is introduced to regularize the displacement field $\delta$ by penalizing large gradients:
\begin{equation}
    \mathcal{L}_{\text{Smooth}} = \sum_{t=1}^{T} \left( \left|\nabla_x \delta_t\right| + \left|\nabla_y \delta_t\right| \right),
\end{equation}
where $\nabla_x \delta_t$ and $\nabla_y \delta_t$ are the gradients of the displacement field along the horizontal and vertical directions, respectively. This loss encourages smoother displacement transitions, preventing unrealistic deformations.

\subsection{Motion and Texture Fusion based Scar Segmentation}

The MTI-MyoScarSeg model leverages the extracted motion information from all frames across the entire cardiac cycle and fuses this with the texture information from the images.
The segmentation network employs a U-Net architecture \cite{ronneberger2015u}. 
It employs the full sequence of images $I_t$ and motion fields $\delta_t$ as input to generate segmentation masks $\hat{S}$ for both LV Myo and scars:
\begin{equation}
    \hat{S} = \mathit{F_{Seg}}(I_t, \delta_t),
\end{equation}
where $\hat{S} \in \mathbb{Z}^{H \times W \times 2}$ represents the predicted binary segmentation masks for Myo and scar.
The network is trained using a combination of Dice loss and binary cross-entropy (BCE) loss:
\begin{align}
    \mathcal{L}_{\text{Seg}} &= \mathcal{L}_\text{Dice}(\hat{S}, S) + \mathcal{L}_\text{BCE}(\hat{S}, S), 
\end{align}
where $S$ denotes the ground-truth segmentation. 

\section{Experiments and Results}
\label{sec:exp_res}

\subsection{Dataset and Evaluation}
\label{ssec:dataset}
We collected paired cine and LGE MRIs dataset of 50 MI patients during 2017-2019. The dataset was randomly divided into 30 training, 2 validation, and 18 testing cases. Both the cine and LGE MRI datasets contain manually segmented Myo labels at the end-diastolic (ED) phase, while the LGE dataset also includes manually labeled scars. The LGE sequences consist of 4 to 11 slices, with an in-plane resolution of 1.33 $\times$ 1.33 mm and a slice spacing of 10 mm. Cine sequences comprise 8 to 11 slices across 25 frames, with an in-plane resolution of 1.77 $\times$ 1.77 mm and a slice spacing of 10 mm.

To obtain the ground truth scar area on the cine MRI, we employ a registration tool \cite{zhuang2018multivariate} to transform the LGE annotations onto cine data.
Specifically, the tool identifies the closest slices along the Z-axis, followed by the application of in-slice rigid and non-rigid registration. 
For evaluation, we employed Dice Score (DS) as a quantitative metric. 

\subsection{Implementation Details}
\label{ssec:Implement}


Cine and LGE images were preprocessed by center-cropping to a resolution of $154 \times 154$ pixels and then resized to $192 \times 192$ pixels. We applied the following data augmentation techniques to improve model generalization: random horizontal and vertical flips, random rotation within 30 degrees in both clockwise and counterclockwise directions, and Gaussian blur.

The motion extraction and segmentation networks were trained using the Adam optimizer over 1000 and 400 epochs, with batch sizes of 16 and 8, respectively. A constant learning rate of $5 \times 10^{-4}$ was applied to both networks.

\subsection{Ablation Study and Comparison Methods}
\begin{table} [t] \center
\caption{Summary of the quantitative evaluation results of different methods.}
\label{table: t1} \small
\begin{tabular}{lcc}
\hline
Method & Dice score \\
\hline
Unet LGE & 0.564 ± 0.163 \\
nn-Unet LGE & 0.631 ± 0.092\\
\hdashline
Baseline & 0.321 ± 0.151\\
Baseline + $\delta_{seg}$ & 0.433 ± 0.111\\
Baseline + $\delta_{seg}$ + $\delta_T$ & 0.502 ± 0.124 \\
Baseline + $\delta_{seg}$ + $\delta_{M}$ & 0.516 ± 0.126 \\
\textbf{Our model} & \textbf{0.593 ± 0.157} \\
\hline
\end{tabular} \\
\vspace{-4mm}
\end{table}

To demonstrate the effectiveness of our model using non-contrast cine MRI, we compared our model with two LGE-based segmentation models: a UNet model with hyper-parameters selected by nn-Unet and a full nn-Unet model.  All experiments are performed with dual task segmentation and the hyper-parameters for proposed model is also adopted by nn-Unet to guarantee a fair comparison.
Furthermore, we performed an ablation study by setting the following configurations:

    
    
    
    
\begin{itemize}
    \item \textbf{Baseline}: Segmentation model using only the ED-frame (single-frame texture) to segment the Myo scar.
    
    \item \textbf{Baseline + Dual Task ($\delta_{\text{Seg}}$)}: Extends the baseline by jointly segmenting the Myo and scar, using only the ED-frame.
    
    \item \textbf{Baseline + Dual Task + Texture ($\delta_T$)}: Incorporates all frames across the cardiac cycle as video input to utilize the implicit temporal information for joint segmentation of the Myo and scar. 
    
    \item \textbf{Baseline + Dual Task + Motion ($\delta_M$)}: Uses the ED-frame combined with extracted motion information from all frames in the cardiac cycle to segment the Myo and scar.
    
    \item \textbf{MTI-MyoScarSeg (Ours)}: Integrates full frame texture information (embedded temporal information) and motion information for comprehensive segmentation of the Myo and scar.
\end{itemize}
\textit{Note:} Temporal information refers to treating the entire cine sequence as video input, expanding the single-frame input to multiple frames across the cardiac cycle. Motion information refers to extracting the displacement field from this video input to capture movement dynamics.

\label{ssec:seg}
\begin{figure*}[htbp]
    \centering
    \includegraphics[width=\textwidth]{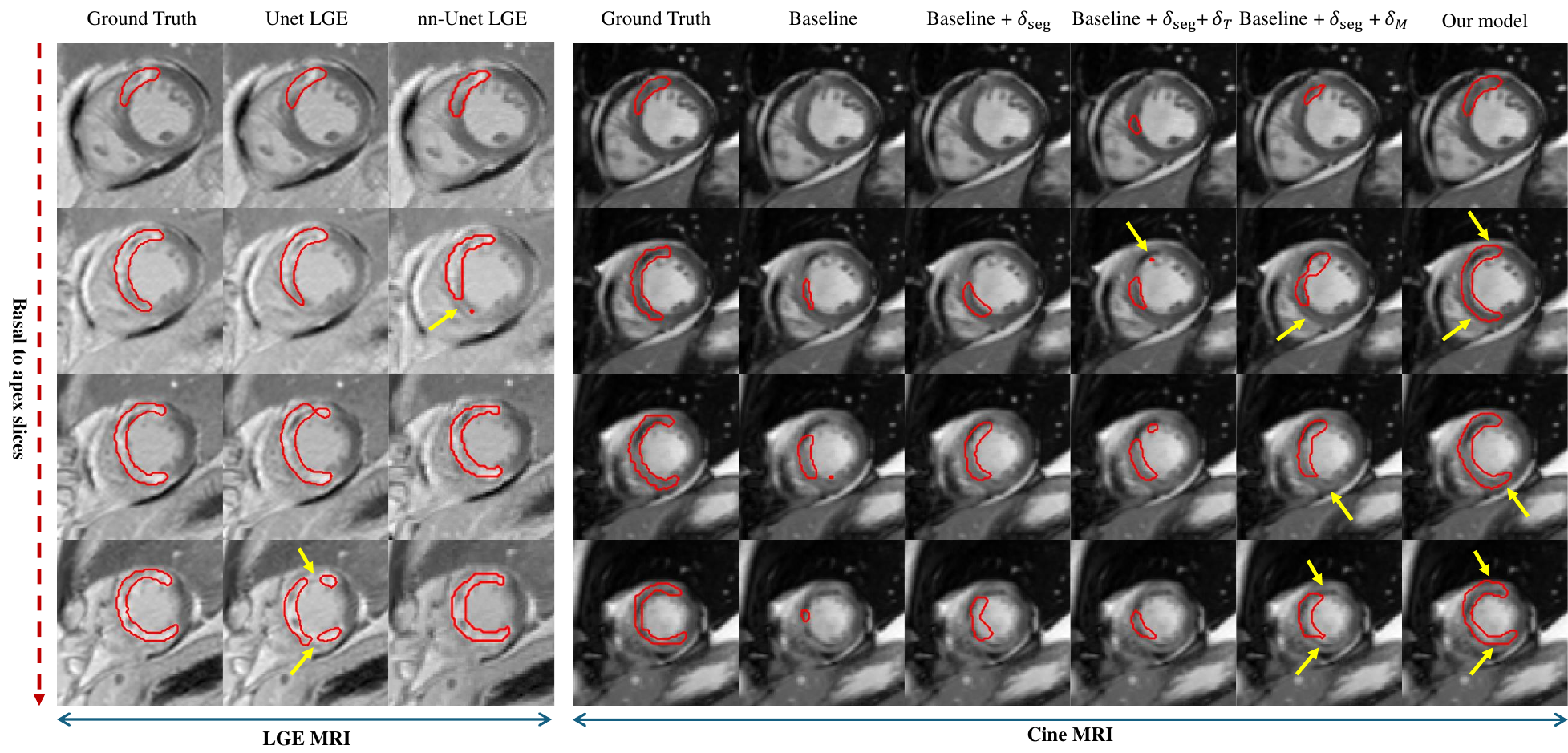}
    \caption{Visualisation of the LV scar segmentation results of different methods from LGE and cine MRIs, respectively.} 
    \label{fig:qualitative}
    \vspace{-4mm}
\end{figure*}

As shown in Table.~\ref{table: t1}, adding Myo segmentation as an auxiliary task ($\delta_{\text{Seg}}$) improved the Dice score by approximately 11\%. Incorporating temporal information from all frames ($\delta_T$) as video input provided an additional 7\% improvement. Including motion features extracted by the motion extraction network ($\delta_M$) further enhanced the performance. Even though our full model only employed non-contrast scans, it outperformed the LGE-based Unet model, demonstrating the effectiveness of integrating motion and texture information.
Although our segmentation results are lower than those achieved by the LGE-based nn-Unet model (which relies on the use of contrast agents) our approach presented the smallest performance gap compared to others. 
Direct comparison between our model and other state-of-the-art results is not possible at this point due to differences in datasets and evaluation methodologies.


Figure \ref{fig:qualitative} presents the qualitative comparison between our model and other segmentation approaches. 
One can see that the proposed model produced consistent and well-defined boundaries across all slices, even compared to LGE-based models. 
As highlighted by the arrows, it achieved sharp and concise boundaries, unlike the nn-Unet, which tended to overestimate the scar length and produce a compressed shape. 
Also,  the proposed model obtained continuous scar segmentation, whereas the LGE-Unet model resulted in fragmented segments.
This might be attributed the introduction of motion information, which can be particularly helpful in regions where intensity changes are subtle and less distinct. 

\subsection{Motion Effectiveness Study}
To validate the effectiveness of our proposed motion extraction approach, we conducted a comparative analysis of different motion estimation methods, as summarized in Table \ref{table:table motion}. Specifically, we evaluated our method against two commonly used techniques: traditional optical flow, implemented with the iterative Lukas-Kanade (ILK) method \cite{zach2007duality}, and frame-to-frame (F2F) registration formulated by DL.

\begin{table}[ht] 
\center
\caption{Quantitative evaluation results for motion estimation methods.}
\label{table:table motion} 
\small
\begin{tabular}{lcc}
\hline
Method & Dice score \\
\hline
Optical flow       & 0.479 ± 0.108 \\
F2F registration   & 0.514 ± 0.121 \\
\textbf{Our model} & \textbf{0.593 ± 0.157} \\
\hline
\end{tabular}
\vspace{-4mm}
\end{table}

The results indicate that the proposed fixed ED frame registration significantly outperformed both ILK and F2F. This is expected as F2F registration may capture only minimal inter-frame motion, potentially resulting in suboptimal guidance for following segmentation model.  In contrast, our method captures cardiac motion dynamics more effectively, leading to improved scar segmentation performance.

\section{Discussion and Conclusion}
\label{sec:discuss}

This study introduced a two-stage model, MTI-MyoScarSeg, that sequentially models cardiac motion and segments Myo scar from non-contrast cine MRI. 
The model directly estimates motion from the images, without relying on segmented Myo contours, either manually labeled or generated by semi-supervised methods.
While this motion may include non-cardiac movements, the incorporation of the texture information from original image data and auxiliary Myo segmentation improved the overall segmentation accuracy.

The approach outperformed single-frame, texture-only, and motion-only methods and achieved superior segmentation performance compared to LGE images under same architecture models. 
Notably, the results demonstrate that the proposed method offers a potential alternative to contrast-enhanced images, with performance approaching that of LGE analyzed by nn-Unet (current upper limit).
However, it also brings the extra computational cost in time for motion extraction. Future work will focus on incorporating inputs from other modalities to conduct multi-modal analysis, validate the algorithm on additional datasets to enhance its generalizability and optimize algorithm to mitigate the time cost.

\section{Acknowledgments}
\label{sec:acknowledgments}

 B. Raman was funded by Wellcome Career Development Award fellowship (302210/Z/23/Z). S. Yang and X. Zhuang were funded by National Natural Science Foundation of China (NSFC) under grant number: 62372115.


\begin{thebibliography}{10}

\bibitem{zeidan2024myocardial}
Rouba~Karen Zeidan and Rita Farah,
\newblock ``Myocardial infarctions in developing countries,''
\newblock in {\em Handbook of Medical and Health Sciences in Developing Countries: Education, Practice, and Research}, pp. 1--30. Springer, 2024.

\bibitem{li2023myops}
Lei Li, Fuping Wu, Sihan Wang, et~al.,
\newblock ``Myo{PS}: A benchmark of myocardial pathology segmentation combining three-sequence cardiac magnetic resonance images,''
\newblock {\em Medical Image Analysis}, vol. 87, pp. 102808, 2023.

\bibitem{bondarenko2005standardizing}
Olga Bondarenko, Aernout~M Beek, Mark~BM Hofman, et~al.,
\newblock ``Standardizing the definition of hyperenhancement in the quantitative assessment of infarct size and myocardial viability using delayed contrast-enhanced cmr,''
\newblock {\em Journal of Cardiovascular Magnetic Resonance}, vol. 7, no. 2, pp. 481--485, 2005.

\bibitem{grani2019comparison}
Christoph Gr{\"a}ni, Christian Eichhorn, Lo{\"\i}c Bi{\`e}re, et~al.,
\newblock ``Comparison of myocardial fibrosis quantification methods by cardiovascular magnetic resonance imaging for risk stratification of patients with suspected myocarditis,''
\newblock {\em Journal of Cardiovascular Magnetic Resonance}, vol. 21, no. 1, pp. 14, 2019.

\bibitem{li2022medical}
Lei Li, Veronika~A Zimmer, Julia~A Schnabel, et~al.,
\newblock ``Medical image analysis on left atrial {LGE MRI} for atrial fibrillation studies: A review,''
\newblock {\em Medical image analysis}, vol. 77, pp. 102360, 2022.

\bibitem{chen2022semi}
Jingkun Chen, Jianguo Zhang, Kurt Debattista, et~al.,
\newblock ``Semi-supervised unpaired medical image segmentation through task-affinity consistency,''
\newblock {\em IEEE Transactions on Medical Imaging}, vol. 42, no. 3, pp. 594--605, 2022.

\bibitem{chen2024dynamic}
Jingkun Chen, Changrui Chen, Wenjian Huang, et~al.,
\newblock ``Dynamic contrastive learning guided by class confidence and confusion degree for medical image segmentation,''
\newblock {\em Pattern Recognition}, vol. 145, pp. 109881, 2024.

\bibitem{xu2020contrast}
Chenchu Xu, Lei Xu, Pavlo Ohorodnyk, Mike Roth, et~al.,
\newblock ``Contrast agent-free synthesis and segmentation of ischemic heart disease images using progressive sequential causal gans,''
\newblock {\em Medical image analysis}, vol. 62, pp. 101668, 2020.

\bibitem{zhang2021toward}
Qiang Zhang, Matthew~K Burrage, et~al.,
\newblock ``Toward replacing late gadolinium enhancement with artificial intelligence virtual native enhancement for gadolinium-free cardiovascular magnetic resonance tissue characterization in hypertrophic cardiomyopathy,''
\newblock {\em Circulation}, vol. 144, no. 8, pp. 589--599, 2021.

\bibitem{xu2018direct}
Chenchu Xu, Lei Xu, Zhifan Gao, et~al.,
\newblock ``Direct delineation of myocardial infarction without contrast agents using a joint motion feature learning architecture,''
\newblock {\em Medical image analysis}, vol. 50, pp. 82--94, 2018.

\bibitem{zhang2019deep}
Nan Zhang, Guang Yang, Zhifan Gao, et~al.,
\newblock ``Deep learning for diagnosis of chronic myocardial infarction on nonenhanced cardiac cine mri,''
\newblock {\em Radiology}, vol. 291, no. 3, pp. 606--617, 2019.

\bibitem{xu2020segmentation}
Chenchu Xu, Joanne Howey, Pavlo Ohorodnyk, et~al.,
\newblock ``Segmentation and quantification of infarction without contrast agents via spatiotemporal generative adversarial learning,''
\newblock {\em Medical image analysis}, vol. 59, pp. 101568, 2020.

\bibitem{alfarano2024estimating}
Andrea Alfarano, Luca Maiano, Lorenzo Papa, et~al.,
\newblock ``Estimating optical flow: A comprehensive review of the state of the art,''
\newblock {\em Computer Vision and Image Understanding}, p. 104160, 2024.

\bibitem{li2024towards}
Lei Li, Julia Camps, Zhinuo Wang, et~al.,
\newblock ``Towards enabling cardiac digital twins of myocardial infarction using deep computational models for inverse inference,''
\newblock {\em IEEE Transactions on Medical Imaging}, 2024.

\bibitem{lian2024frequency}
Shichang Lian, Zhifan Gao, Hui Wang, et~al.,
\newblock ``Frequency-enhanced geometric-constrained reconstruction for localizing myocardial infarction in 12-lead electrocardiograms,''
\newblock {\em IEEE Transactions on Biomedical Engineering}, 2024.

\bibitem{ronneberger2015u}
Olaf Ronneberger, Philipp Fischer, et~al.,
\newblock ``U-net: Convolutional networks for biomedical image segmentation,''
\newblock in {\em Medical image computing and computer-assisted intervention}. Springer, 2015, pp. 234--241.

\bibitem{jaderberg2015spatial}
Max Jaderberg, Karen Simonyan, Andrew Zisserman, et~al.,
\newblock ``Spatial transformer networks,''
\newblock {\em Advances in neural information processing systems}, vol. 28, 2015.

\bibitem{zhuang2018multivariate}
Xiahai Zhuang,
\newblock ``Multivariate mixture model for myocardial segmentation combining multi-source images,''
\newblock {\em IEEE transactions on pattern analysis and machine intelligence}, vol. 41, no. 12, pp. 2933--2946, 2018.

\bibitem{zach2007duality}
Christopher Zach, Thomas Pock, and Horst Bischof,
\newblock ``A duality based approach for realtime tv-l 1 optical flow,''
\newblock in {\em Pattern Recognition: 29th DAGM Symposium, Heidelberg, Germany, September 12-14, 2007. Proceedings 29}. Springer, 2007, pp. 214--223.

\end{thebibliography}

\end{document}